\documentclass[twocolumn,english,aps,prl,superscriptaddress,nofootinbib]{revtex4-1}
\pdfoutput=1
\usepackage{graphicx}
\usepackage{amssymb,amsmath}
\usepackage{textcomp} 
\usepackage[bold]{hhtensor}
\usepackage{natbib}
\usepackage{lipsum}
\usepackage{float}
\newcommand\blfootnote[1]{%
  \begingroup
  \renewcommand\thefootnote{}\footnote{#1}%
  \addtocounter{footnote}{-1}%
  \endgroup
}

\newcommand{\tocless}[2]{\bgroup\let\addcontentsline=\nocontentsline#1{#2}\egroup}

\begin{document}

\title{Monolithic Ultrahigh-Q Lithium Niobate Microring Resonator}
\author{Mian Zhang*}
\affiliation{John A. Paulson School of Engineering and Applied Sciences, Harvard University, Cambridge, Massachusetts 02138, USA}
\author{Cheng Wang*}
\blfootnote{\vskip -0.2in *These authors contributed equally to this work}
\affiliation{John A. Paulson School of Engineering and Applied Sciences, Harvard University, Cambridge, Massachusetts 02138, USA}
\author{Rebecca Cheng}
\affiliation{John A. Paulson School of Engineering and Applied Sciences, Harvard University, Cambridge, Massachusetts 02138, USA}
\affiliation{Department of Physics, Brown University, Providence, RI 02912, USA}
\author{Amirhassan Shams-Ansari}
\affiliation{John A. Paulson School of Engineering and Applied Sciences, Harvard University, Cambridge, Massachusetts 02138, USA}
\affiliation{Department of Electrical Engineering and Computer Science, Howard University, Washington DC, USA}
\author{Marko Loncar}
\affiliation{John A. Paulson School of Engineering and Applied Sciences, Harvard University, Cambridge, Massachusetts 02138, USA}
\date{\today}

\begin{abstract}
We demonstrate an ultralow loss monolithic integrated lithium niobate photonic platform consisting of dry-etched subwavelength waveguides. We show microring resonators with a quality factor of 10$^7$ and waveguides with propagation loss as low as 2.7 dB/m. 
\end{abstract}
\maketitle

\date{\today}
\newcommand{\nocontentsline}[3]{}

Lithium niobate (LN) is a material with wide range of applications in optical and microwave technologies, owing to its unique properties that include large second order nonlinear susceptibility ($\chi^{(2)}$= 30 pm/V), large piezoelectric response ($C_{33} \sim$ 250 C/m$^2$), wide optical transparency window (350 nm $-$ 5 $\mu$m) and high refractive index ($\sim$ 2.2) \cite{maleki_2}. Conventional LN components, including fiber-optic modulators and periodically poled frequency converters have been the workhorse of the optoelectronic industry. The performances of these components have the potential to be dramatically improved as optical waveguides in bulk LN crystals are defined by ion-diffusion or proton-exchange methods which result in low index contrast and weak optical confinement. Integrated LN platform, featuring sub-wavelength scale light confinement and dense integration of optical and electrical components, has the potential to revolutionize optical communication and microwave photonics  \cite{maleki_2,chen_hybrid_2014,rao_heterogeneous_2015,chang_thin_2016,luo_-chip_2017,wang_high-q_2015,wang_nanophotonic_2017}.

The major road-block for practical applications of integrated LN photonics is the difficulty of fabricating devices that simultaneously achieve low optical propagation loss and high confinement. Recently developed thin-film LN-on-insulator technology makes this possible, and has resulted in the development of two complementary approaches to define nanoscale optical waveguides: hybrid and monolithic. The hybrid approach integrates an easy-to-etch material (e.g. silicon or silicon nitride) with LN thin films to guide light \cite{chen_hybrid_2014,rao_heterogeneous_2015,chang_thin_2016} with a relatively low propagation loss (0.3 dB/cm) \cite{chang_thin_2016}. However, the resulting optical modes only partially reside in the active material region (i.e. LN), reducing the nonlinear interaction efficiency. The monolithic approach relies on direct etching of LN to achieve high optical confinement in the active region, but has suffered from a relatively high propagation loss. While freestanding LN microdisk resonators have achieved optical quality factors ($Q$) of $\sim 10^6$ \cite{luo_-chip_2017,wang_high-q_2015}, integrated microring resonators typically feature $Q \sim 10^5$ with waveguide propagation loss on the order of 3 dB/cm \cite{wang_nanophotonic_2017}. Since LN is perceived as a difficult-to-etch material, it is commonly accepted that low-loss propagation in a monolithic waveguide is not possible, and that therefore it is not the most promising path forward.

\begin{figure}
	\centering
	\includegraphics[angle=0,width=0.45\textwidth]{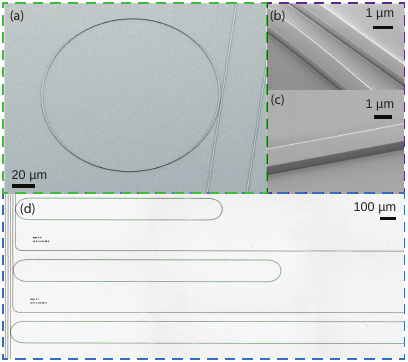}
	
	\caption{\label{fig1}\textbf{Fabricated Devices} Scanning electron microscope (top) and optical microscope (bottom) images of a microring and micro-racetrack resonators of various length. (a) A microring resonator. (b) Zoom-in of the coupling region. (c) Close-up of the etched waveguide. (d) Fabricated racetrack resonators.}
		
	\end{figure}
	
	\begin{figure}
		\centering
		\includegraphics[angle=0,width=0.4\textwidth]{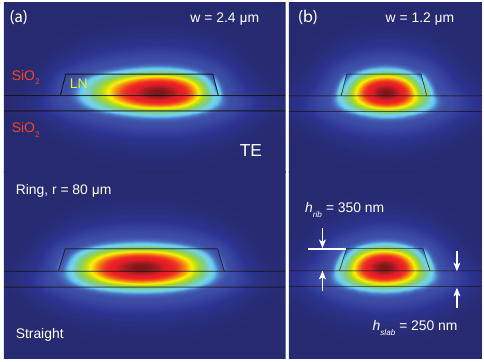}
		
		\caption{\label{fig2}\textbf{Simulated optical modes} Numerical finite element simulation (COMSOL) of the optical modes in straight and bent waveguides with width (a) 2.4 $\mu$m and (b) 1.2 $\mu$m.}
			
		\end{figure}
		
		\begin{figure*}[ht]
			\centering
			\includegraphics[angle=0,width=0.8\textwidth]{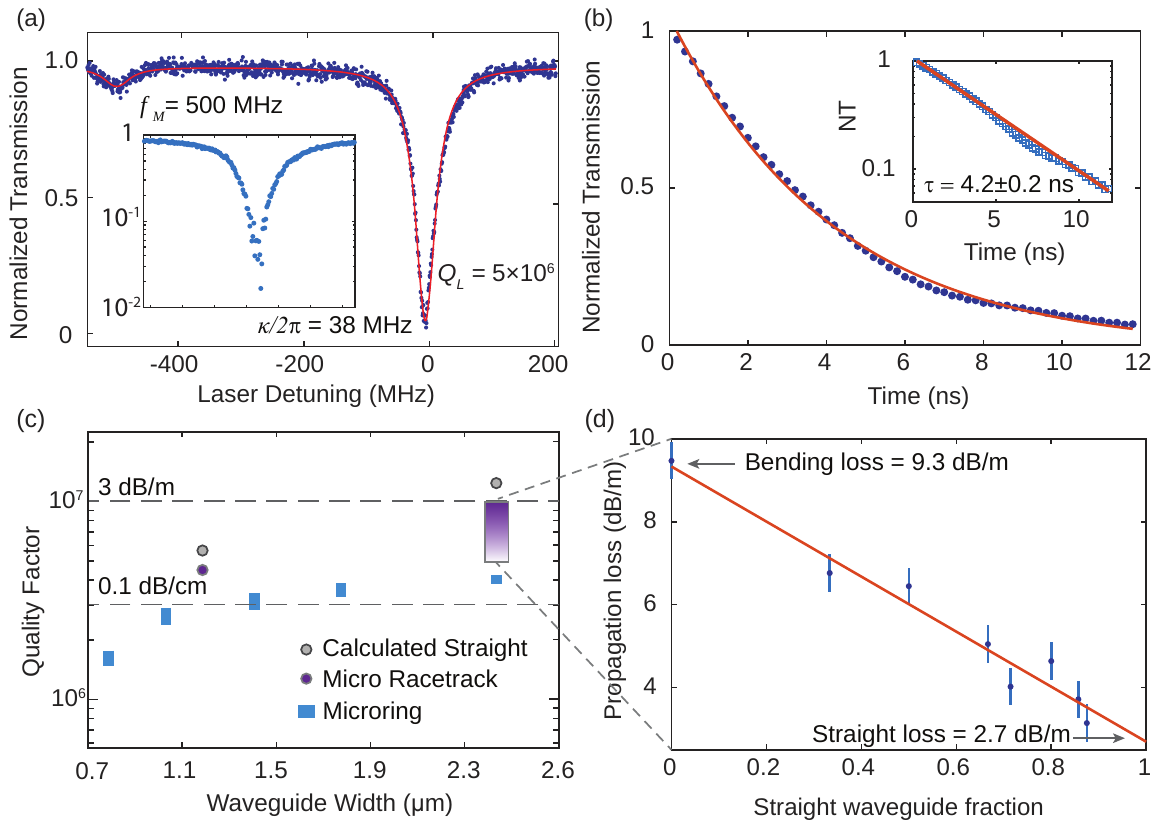}
			
			\caption{\label{fig3}\textbf{Optical Characterization} (a) Highest measured $Q$ factor of a micro-racetrack resonator exhibiting a loaded $Q = 5\times 10^6$. The laser is also modulated by a precise frequency source at 500 MHz for calibration. Inset: the resonator is close to critical coupling. (b) Ring-down measurement of the same device. (c) Measured quality factors for different resonators. (d) Extracted propagation loss for racetrack resonators. All the measurements are conducted around 1590 nm.}
			
		\end{figure*}
		
		Here we challenge the \emph{status quo} and show that sub-wavelength scale lithium niobate waveguides can be fabricated with a propagation loss as low as 2.7 $\pm$ 0.3 dB/m through an optimized etching process. We also demonstrate ultra-high $Q$ factor optical cavities with intrinsic $Q = 10^7$.
		
		We fabricate waveguide coupled microring and racetrack resonators with a bending radius $r$ = 80 $\mu m$, and various straight arm lengths $l$ and waveguide widths $w$ (Fig. \ref{fig1}). The optical modes in these resonators have different interaction strength with the etched sidewalls which allows us to identify the source of the optical losses (Fig. \ref{fig2}). We fabricate the devices using a 600 nm thick X-cut LN thin film on 2 $\mu$m of silicon dioxide (SiO$_2$) on a silicon substrate (NanoLN). We use standard electron beam lithography to define patterns in hydrogen silsesquioxane (HSQ) resist with multi-pass exposure. The patterns are subsequently transferred into the LN thin film using a commercial inductively coupled plasma reactive ion etching (ICP RIE) tool. We use Ar+ plasma to physically etch LN where the plasma power and chamber condition are tuned to minimize surface redeposition of removed LN and other contaminations present in the chamber. We etch a total of $h_{\textrm{rib}}$ = 350 nm of LN with a bias power of 112 W leaving a $h_{\textrm{slab}}$ = 250 nm thin LN slab. After cleaning and removal of the resist, the devices are cladded by depositing 1.5 $\mu$m of SiO$_2$ using plasma-enhanced chemical vapor deposition (PECVD).

	The optical $Q$ factors of these resonators are measured using a tunable telecom external cavity diode laser (Santec TSL-510). To calibrate the $Q$ measurement, we also modulate (500 MHz) the laser output using an external electro-optic modulator. Fig. \ref{fig3}a shows a critically-coupled micro-racetrack resonator with a loaded (intrinsic) $Q$ factor of 5.0 (10.0$\pm 0.7$) million at an optical excitation wavelength of 1590 nm. To confirm the $Q$ factor, we directly measure the photon lifetime of the resonator using ring-down measurement (Fig. \ref{fig3}b). The fitted data show a lifetime of 4.2$\pm$0.2 ns in good agreement with the spectral measurements.  
	We show that sidewall scattering is still a significant loss channel in our devices by investigating structures with different waveguide widths. Fig \ref{fig3}c shows the intrinsic $Q$ measured for fabricated ring and racetrack resonators. Increasing waveguide width from 800 nm to 2.4 $\mu$m leads to a $Q$ improvement from $\sim 1.5$ to $\sim 4$ million. This improvement is expected from the reduced interaction of the light and the etched sidewalls (Fig. \ref{fig2}).
	
	We extract the straight optical waveguide propagation loss by comparing the optical $Q$ factors of micro-racetrack resonators with different straight section lengths from 0 to 14 mm, with the same waveguide width of 2.4 $\mu$m and bending radius of 80 $\mu$m. Indeed, racetrack resonators with longer straight sections exhibit higher optical $Q$ factors as light in the straight section interact less with the sidewall scattering centers than in the bending region (Fig \ref{fig2}). The results are shown in Fig. \ref{fig3}d, where we extract a straight waveguide loss of 2.7$\pm$0.3 dB/m from the intersection of the fitted line. The extracted bending loss of 9.3$\pm$0.9 dB/m agrees well with the measurements in ring resonators. By comparing the straight section loss of two race-track resonators with different widths (1.2 and 2.4 $\mu$m) and using numerical modeling, we estimate an upper bound of 1.5 dB/m for the intrinsic loss of our platform. The possible contributing factors are ion-implantation damage introduced during the LN thin-film production process, top-surface roughness due to LN polishing, and finite PECVD SiO$_2$ cladding absorption. Importantly, these loss mechanisms could be mitigated by a combination of defects annealing and finer top-surface polishing \cite{ji_ultra-low-loss_2017}. We expected that the these loss contributions of the monolithic LN platform could be further reduced towards its material intrinsic loss limit of $<$ 0.1 dB/m \cite{maleki_2}. 
	
	In conclusion, we show that the monolithic LN nanophotonic platform is a viable path forward. The ultralow loss, high optical confinement and tight bending radius combined with the ability to integrate microwave electrodes \cite{wang_nanophotonic_2017} will bring electro-optic \cite{wang_nanophotonic_2017,chen_hybrid_2014} nonlinear optical \cite{chang_thin_2016} systems into a new design parameter space that has been inaccessible so far. This could enable a wide range of applications including those in ultralow loss quantum photonics \cite{obrien_optical_2007} ,coherent microwave to optical conversion \cite{javerzac-galy_-chip_2016,soltani_efficient_2017} and active topological photonics \cite{lin_photonic_2016}. We emphasize that the LN device layer sits atop a standard silicon handle wafer and therefore our platform can also be integrated with many existing photonic technologies.  Finally, we hope that our work will stimulate further activities and a renewed interest of the monolithic LN nanophotonic approach, that could ultimately result in the development of dedicated LN nanofabrication foundries and/or introduction of etched LN into existing silicon photonics processes. 
	
	The authors thank National Science Foundation ECCS-1609549, DMR-1231319) and REU program; Harvard University Office of Technology Development Physical Sciences and Engineering Accelerator; ARL Center for Distributed Quantum Information W911NF1520067. Device fabrication was performed at the center for nanoscale systems (CNS) at Harvard University. 

\bibliography{reference}
\end{document}